\documentclass[12pt,preprint]{aastex}
\usepackage{epsfig}
\usepackage{rotating}
\usepackage{natbib}
\def\etal{{et al.\thinspace}}

\def\spose#1{\hbox to 0pt{#1\hss}}

\def\multleft#1{\hbox to size{\vbox {\halign {\lft{##}\cr #1}}\hfill}\par}
\def\multright#1{\hbox to size{\vbox {\halign {\rt{##}\cr #1}}\hfill}\par}

\def\degmark{^\circ}
\def\boxit#1{\vbox{\hrule\hbox{\vrule\kern3pt\vbox{\kern3pt
          #1 \kern3pt}\kern3pt\vrule}\hrule}}

\def\cm{{\rm\thinspace cm}}

\def\erg{{\rm\thinspace erg}}
\def\eV{{\rm\thinspace eV}}

\def\keV{{\rm\thinspace keV}}
\def\km{{\rm\thinspace km}}
\def\kpc{{\rm\thinspace kpc}}

\def\Mpc{{\rm\thinspace Mpc}}

\def\ph{{\rm\thinspace ph}}
\def\s{{\rm\thinspace s}}
\def\ks{{\rm\thinspace ks}}

\def\cts{{\rm\thinspace cts}}

\def\ergcmps{\hbox{$\erg\cm\s^{-1}\,$}}

\def\ergpcmsqps{\hbox{$\erg\cm^{-2}\s^{-1}\,$}}

\def\ergps{\hbox{$\erg\s^{-1}\,$}}

\def\kmps{\hbox{$\km\s^{-1}\,$}}

\def\pcmsq{\hbox{$\cm^{-2}\,$}}

\def\phpcmsqps{\hbox{$\ph\cm^{-2}\s^{-1}\,$}}

\def\kmpspMpc{\hbox{$\kmps\Mpc^{-1}$}}

\def\ctsps{\hbox{$\cts\s^{-1}$}}

\makeatletter

\let\@internalcite\cite
\def\cite{\@ifstar{\citey}{\citefull}}
\def\citefull{\def\astroncite##1##2{##1\ ##2}\@internalcite}
\def\citey{\def\astroncite##1##2{##1\ (##2)}\@internalcite}
\def\citeyear{\def\astroncite##1##2{##2}\@internalcite}
\def\citename{\def\astroncite##1##2{##1}\@internalcite}
\def\@citex[#1]#2{\if@filesw\immediate\write\@auxout{\string\citation{#2}}\fi
  \def\@citea{}\@cite{\@for\@citeb:=#2\do
    {\@citea\def\@citea{; }\@ifundefined
       {b@\@citeb}{{\bf ??}\@warning
       {Citation `\@citeb' on page \thepage \space undefined}}%
{\csname b@\@citeb\endcsname}}}{#1}}

\def\@cite#1#2{#1\if@tempswa #2\fi}
\def\@biblabel#1{}

\def\astroncite#1#2{#1\ #2}
\makeatother

\begin{document}

\title{The Spin of the Supermassive Black Hole in NGC~3783}

\author{L.~W.~Brenneman\altaffilmark{1}, 
C.~S.~Reynolds\altaffilmark{2,3},
M.~A.~Nowak\altaffilmark{4},
R.~C.~Reis\altaffilmark{5},
M.~Trippe\altaffilmark{2},
A.~C.~Fabian\altaffilmark{5},
K.~Iwasawa\altaffilmark{6},
J.~C.~Lee\altaffilmark{7},
J.~M.~Miller\altaffilmark{8},
R.~F.~Mushotzky\altaffilmark{2,3},
K.~Nandra\altaffilmark{9},
M.~Volonteri\altaffilmark{8}}
\altaffiltext{1}{Harvard-Smithsonian CfA, 60 Garden St. MS-67, Cambridge, MA~02138~USA}
\altaffiltext{2}{Dept. of Astronomy, University of Maryland, College
  Park, MD~20742~USA}
\altaffiltext{3}{Joint Space Science Institute (JSI), University of Maryland, College
  Park, MD~20742~USA}
\altaffiltext{4}{MIT Kavli Institute for Astrophysics, Cambridge, MA~02139~USA}
\altaffiltext{5}{Institute of Astronomy, University of Cambridge, Madingley Rd., Cambridge CB3 0HA, UK}
\altaffiltext{6}{Universitat de Barcelona}
\altaffiltext{7}{Dept. of Astronomy, Harvard University,
  Harvard-Smithsonian CfA, 60 Garden St. MS-6, Cambridge, MA~02138~USA}
\altaffiltext{8}{Dept. of Astronomy, University of Michigan, Ann Arbor, Michigan~48109~USA}
\altaffiltext{9}{Max-Planck-Institut fu?r Extraterrestrische Physik, Giessenbachstrasse 1, 85740 Garching, Germany}

\begin{abstract}
\noindent The {\it Suzaku AGN Spin Survey} is designed to determine
the supermassive black hole spin in six nearby active galactic nuclei
(AGN) via deep {\it Suzaku} stares, thereby giving us our first
glimpse of the local black hole spin distribution.  Here, we present
an analysis of the first target to be studied under the auspices of
this Key Project, the Seyfert galaxy NGC~3783.  Despite complexity in
the spectrum arising from a multi-component warm absorber, we detect
and study relativistic reflection from the inner accretion disk.
Assuming that the X-ray reflection is from the surface of a flat disk around a
Kerr black hole, and that no X-ray reflection occurs within the general
relativistic radius of marginal stability, we determine a lower limit on the
black hole spin of $a \geq 0.88$ (99\% confidence).
We examine the robustness of this result to the assumption of the analysis, and
present a brief discussion of spin-related selection biases that might affect
flux-limited samples of AGN.
\end{abstract}

\section{Introduction}
\label{sec:intro}

Ever since the seminal work of \citet{Penrose1969} and \citet{BZ1977},
it has been recognized that black hole spin may be an
important source of energy in astrophysics.  Of particular note is the
role that black hole spin may play in relativistic jets such as those
seen in radio-loud active galactic nuclei (AGN) --- the magnetic
extraction of the rotational energy of a rapidly spinning black hole
is the leading contender for the fundamental energy source of such
jets.  Indeed, it has been suggested that the spin of the central
supermassive black hole (SMBH) is a crucial parameter in determining
whether an AGN can form powerful jets (i.e., whether the source is
radio-quiet or radio-loud; \citealt{Wilson1995}), although the
accretion rate/mode must clearly have a role to play \citep{Sikora2007}.

However, the importance of black hole spin goes beyond its role as a
possible power source.  The spin distribution of the SMBH population
(and its dependence on SMBH mass) encodes the black hole growth
history \citep{Moderski1996,Volonteri2005,Berti2008}.  
In essence, if local SMBHs have obtained most of their mass during prolonged
prograde accretion events in a quasar phase of activity, or in major mergers 
with similar mass SMBHs, we would expect a population of rapidly rotating SMBHs 
($a>0.6$) due to the angular momentum accreted
from the disks or transferred at merger \citep{Rezzolla2008}.  Here we define $a
\equiv cJ/GM^2$, where $J$ is the angular momentum of the black hole and $M$ is
its mass.  On the other
hand, if mergers with much smaller SMBHs \citep{Hughes2003} or randomly-oriented
accretion events of small packets of material \citep{King2007} have been the
dominant growth mechanism, most of the SMBHs would be spinning at a much more
modest rate.

To date, the cleanest probe of strong gravitational physics around
SMBHs, including the effects of black hole spin, comes from examining
relativistically-broadened spectral features that are emitted from the
surface layers of the inner accretion disk in response to irradiation
by the hard X-ray source \citep{ReynoldsNowak2003,Miller2007}.  
These spectral features have been observed and well-characterized in both
AGN \citep{Tanaka1995,Fabian1995} and stellar-mass black hole
systems \citep{Miller2002,Reis2008}. The strongest
feature in this so-called ``reflection spectrum'' is the fluorescent
Fe K$\alpha$ line (rest frame energy of $6.4 \keV$); in contrast to
lines from other elements, its relative abundance, high energy and
fluorescent yield make Fe K$\alpha$ visible above the typical
power-law continuum seen commonly in BH systems. Extreme Doppler
effects and gravitational redshifts combine to give this line (and all
other features in the reflection spectrum) a characteristic broadened
and skewed profile \citep{Fabian1989,Laor1991}.   Modern high
signal-to-noise (S/N) datasets from {\it XMM-Newton} and {\it Suzaku},
combined with the latest models of reflection from an ionized
accretion disk \citep[e.g.,][]{Ross2005} and variable-spin
relativistic smearing models \citep[e.g.,][]{Brenneman2006,Dauser2010}, are
giving us our first glimpses at the spins of SMBHs.   
However, due to
the high S/N required to characterize the subtle effects of SMBH spin,
interesting spin constraints have only been determined for a small
handful of AGN at present
(MCG--6-30-15, \citealt{Brenneman2006}; Fairall 9,
\citealt{Schmoll2009}; SWIFT~J2127.4$+$5654, \citealt{Miniutti2009}; 1H0707--495, 
\citealt{Zoghbi2010}; Mrk~79, \citealt{Gallo2011}; Mrk~335 \& NGC~7469
\citealt{Patrick2010}; see Table~2).

Under the auspices of the {\it Suzaku} Key Projects program, we have
initiated a series of deep quasi-continuous observations of bright,
nearby AGN with the purpose of characterizing relativistic disk
features in the spectra and setting constraints on the SMBH spin
({\it Suzaku AGN Spin Survey; PI C.~Reynolds}).  In this {\it
  Paper}, we present results from the first object to be studied
under this program, the Seyfert 1.5 galaxy NGC~3783 ($z=0.00973$;
\citealt{Theureau1998}).   This object possesses a high-column density 
and multi-component warm absorber that has been well-studied by every
spectroscopic X-ray observatory, including a $900 \ks$ campaign by
{\it Chandra} using the High-Energy Transmission Grating Spectrometer
\citep[HETGS;][]{Kaspi2002,Krongold2003,Netzer2003}.   
We show that, despite the presence of this complex warm 
absorber, reflection signatures from the inner accretion disk can
be identified and characterized with sufficient accuracy to constrain
SMBH spin.  We conclude that the SMBH is rapidly spinning with
$a>0.93$ ($90\%$ confidence).  This result is shown to be robust to
the exclusion of the complex, soft region of the X-ray spectrum as well as 
to uncertainties in the XIS/PIN cross-normalization.  

This paper is organized as follows.  Section~\ref{sec:obs} discusses the
{\it Suzaku} observation of NGC~3783 and the basic reduction of the data.
Section~3 then presents our modeling of the $0.7-45 \keV$ time-averaged
spectrum of NGC~3783, including our newly-derived constraints on the
SMBH spin.  Section~4 summarizes our conclusions on NGC~3783 and 
addresses the role of spin-dependent selection biases in AGN samples.

\section{Observations and Data Reduction}
\label{sec:obs}

NGC~3783 was observed by {\it Suzaku} quasi-continuously for the
period 10--15 July 2009, with the source placed in the Hard X-ray
Detector (HXD) nominal aimpoint.   After eliminating Earth
occultations, South Atlantic Anomaly (SAA) passages, and other high
background periods, the observation contains 210\,ks of ``good"
on-source exposure.  The XIS data (XIS~0, XIS~1 and XIS~3; XIS~2 has
been inoperable since November 2006) were reprocessed using the {\tt
  xispi} script in accordance with the {\it Suzaku} ABC
Guide\footnote{http://heasarc.gsfc.nasa.gov/docs/suzaku/analysis/abc/}
along with the latest version of the CALDB (as of 29 March, 2010).  
XIS spectra and light curves were then produced according to the
procedure outlined in the ABC Guide.  For the XIS spectra, we combined
data from the front-illuminated (FI) detectors XIS~0+3 data using the
{\tt addascaspec} script in order to increase S/N.  The
XIS spectra, responses and backgrounds were then rebinned to $512$
spectral channels from the original $4096$ in order to speed up
spectral model fitting without compromising the resolution of the
detectors.  Finally, the XIS spectra were grouped to a minimum of $25$
counts per bin in order to facilitate robust $\chi^2$ fitting.
The merged, background-subtracted, time-averaged FI spectrum has a net
count rate of $4.960 \pm 0.002 \ctsps$ for a total of $1.04\times 10^6$
counts.  The total number of $2-10 \keV$ counts is $6.26\times 10^5$.
The total XIS~1 count rate is $3.043\pm 0.004 \ctsps$ for a total of
$6.40\times 10^5$, or $3.14\times 10^5$ when restricting to $2-10
\keV$.   For all of the fitting presented in this paper, we allow for a 
global flux cross-normalization error between the FI and XIS~1 spectra.
The XIS~1/FI cross-normalization is allowed to be a free parameter, and is 
found to be approximately 1.03.

The HXD/PIN instrument detected NGC~3783, though the GSO did not.
Data from PIN were again reduced as per the {\it Suzaku} ABC Guide.
For background subtraction, we used the ``tuned'' non X-ray background
(NXB) event file for July 2009 from the {\it Suzaku} CALDB, along with
the appropriate response file and flat field file for epoch 5 data.
The NXB background contributed a count rate of $0.2165 \pm 0.0004\ctsps$ to the total X-ray
background from $16-45 \keV$.
We modeled the cosmic X-ray background (CXB) contribution as per the
ABC Guide, simulating its spectrum in XSPEC \citep{Arnaud1996}.
The simulated CXB spectrum contributed a count rate of $0.0237 \pm
0.0002\ctsps$ to the total X-ray background from $16-45 \keV$.  The NXB and
CXB files were combined to form a single PIN background spectrum.   
In comparison, the PIN data had a count rate of $0.4561 \pm 0.0027\ctsps$ over the same
energy range, roughly twice that of the total background.

Because the PIN data only contain 256 spectral channels (vs. the 4096
channels in the unbinned XIS data), rebinning to 25 counts per bin was
not necessary in
order to facilitate $\chi^2$ fitting.  Rather, we rebinned the PIN
spectrum to have a S/N of 5 in each energy bin,
which limited our energy
range to $16-45 \keV$.  After reduction, filtering and background
subtraction, the PIN spectrum had a net $16-45 \keV$ count rate of $0.360
\pm 0.002 \ctsps$.  We also added $3\%$ systematic errors to the PIN
data to account for the uncertainty in the non-X-ray background data
supplied by the {\it Suzaku} calibration team.
For most of the spectral fitting presented in this paper,
we assume a PIN/XIS-FI cross-normalization factor of 1.18 as per the 
{\it Suzaku} memo
2008-06\footnote{http://heasarc.gsfc.nasa.gov/docs/suzaku/analysis/watchout.html}.
However, for the final fits used to constrain the black hole spin in \S\ref{sec:spin},
we investigate the effect of allowing this cross-normalization to be a free
parameter.

\begin{figure}[t]
\begin{center}
\includegraphics[width=0.7\textwidth,angle=270]{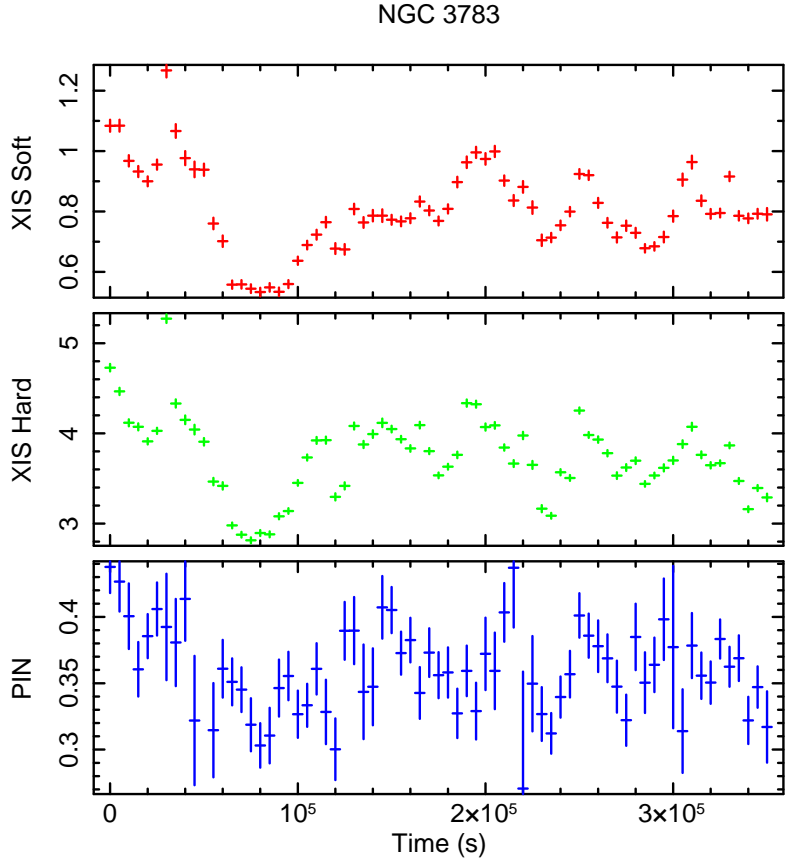}
\end{center}
\caption{{\small Co-added and background subtracted XIS
    light curves in the soft ($0.3-1 \keV$) and hard ($2-10 \keV$) bands,
    together with the background subtracted PIN ($16-45 \keV$) light curve.  These
    light curves are shown with $5000 \s$ bins.}}
\label{fig:lc}
\end{figure}

XIS light curves (both hard and soft band) as well as PIN light curves
are shown in Fig.~\ref{fig:lc}.   In the soft band ($0.3-1 \keV$), the
source is observed to undergo variability by a factor of almost two.
Most of the large amplitude variability occurs on timescales of
$50-100 \ks$, although there are occasional sharp flares/dips that
occur much more rapidly.   Also noteworthy is that the amplitude of
variability decreases as one considers higher-energy bands, suggesting
``pivoting" of the spectrum about some energy above the {\it
  Suzaku}/PIN band.   The detailed nature of this spectral variability
will be the subject of another publication (Reis \etal, in
preparation).   
For the remainder of this paper, we examine the high
S/N spectrum from the time-averaged dataset.  We restrict our energy range to
$0.7-10 \keV$ in the XIS data, ignoring energies below $0.7 \keV$ and from
$1.5-2.5 \keV$ to avoid areas of significant deviation between the three
detectors, i.e., regions of known calibration uncertainty.

\section{Analysis of the time-averaged {\it Suzaku} spectrum}
\label{sec:analysis}

\subsection{A First Look at the Hard-band Spectrum}
\label{sec:hard}

\begin{figure}[t]
\hbox{
\includegraphics[width=0.33\textwidth,angle=270]{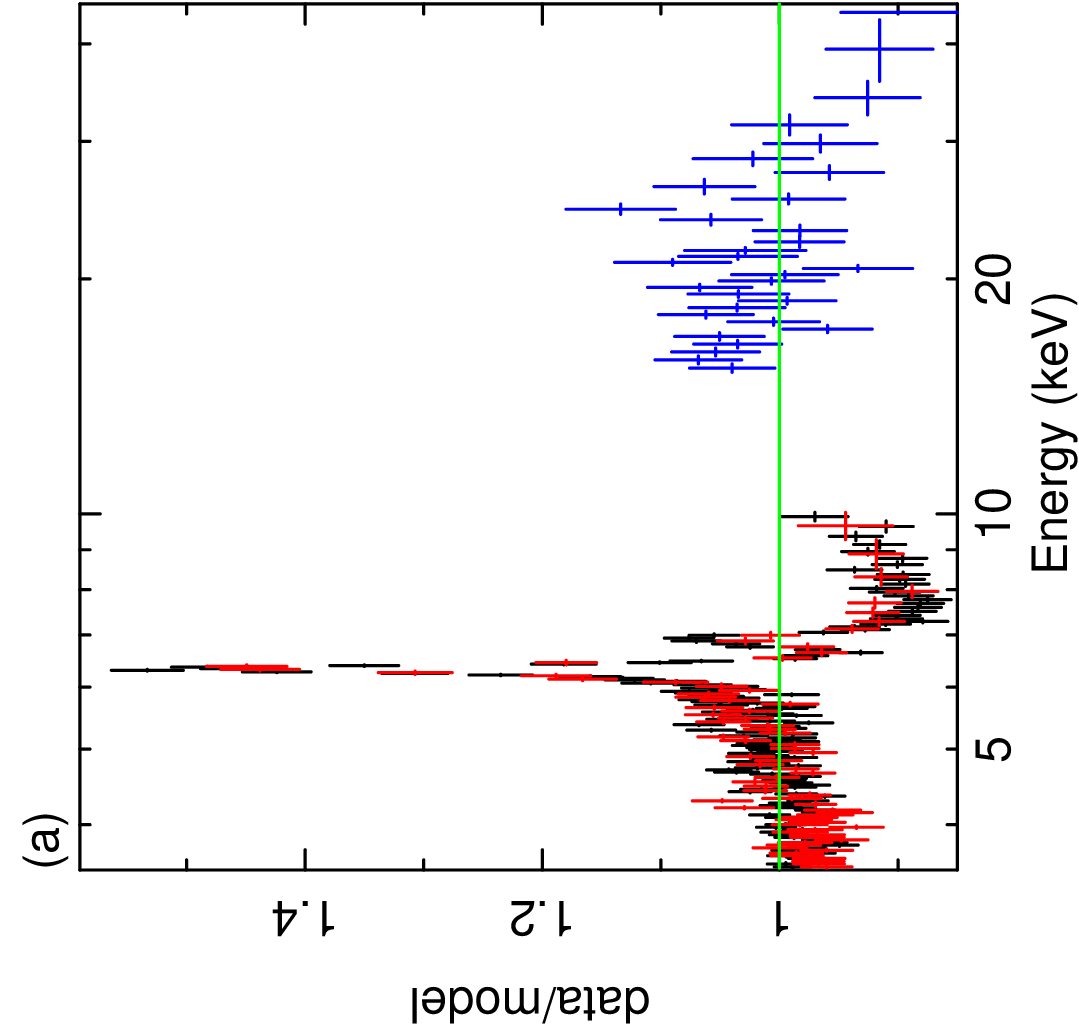}
\includegraphics[width=0.33\textwidth,angle=270]{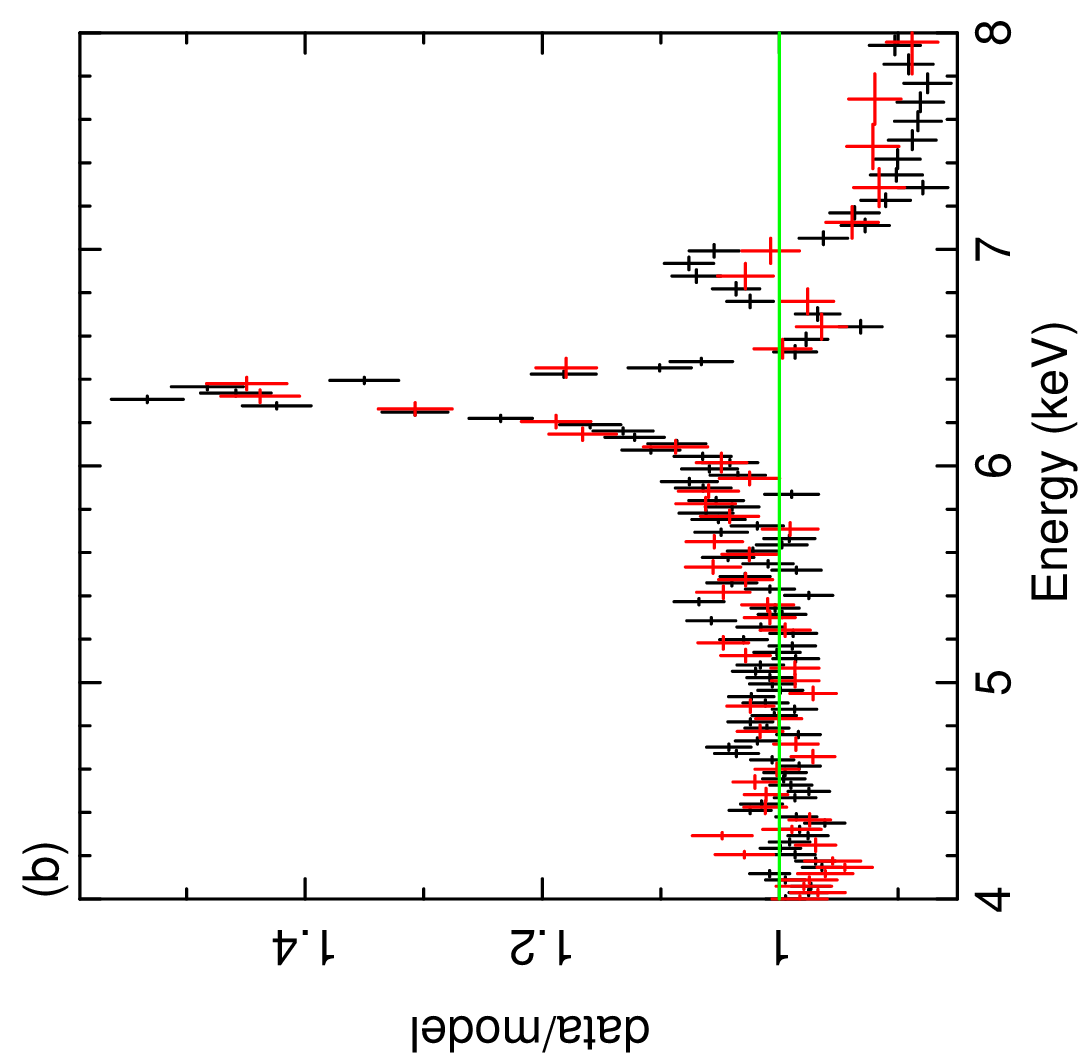}
\includegraphics[width=0.33\textwidth,angle=270]{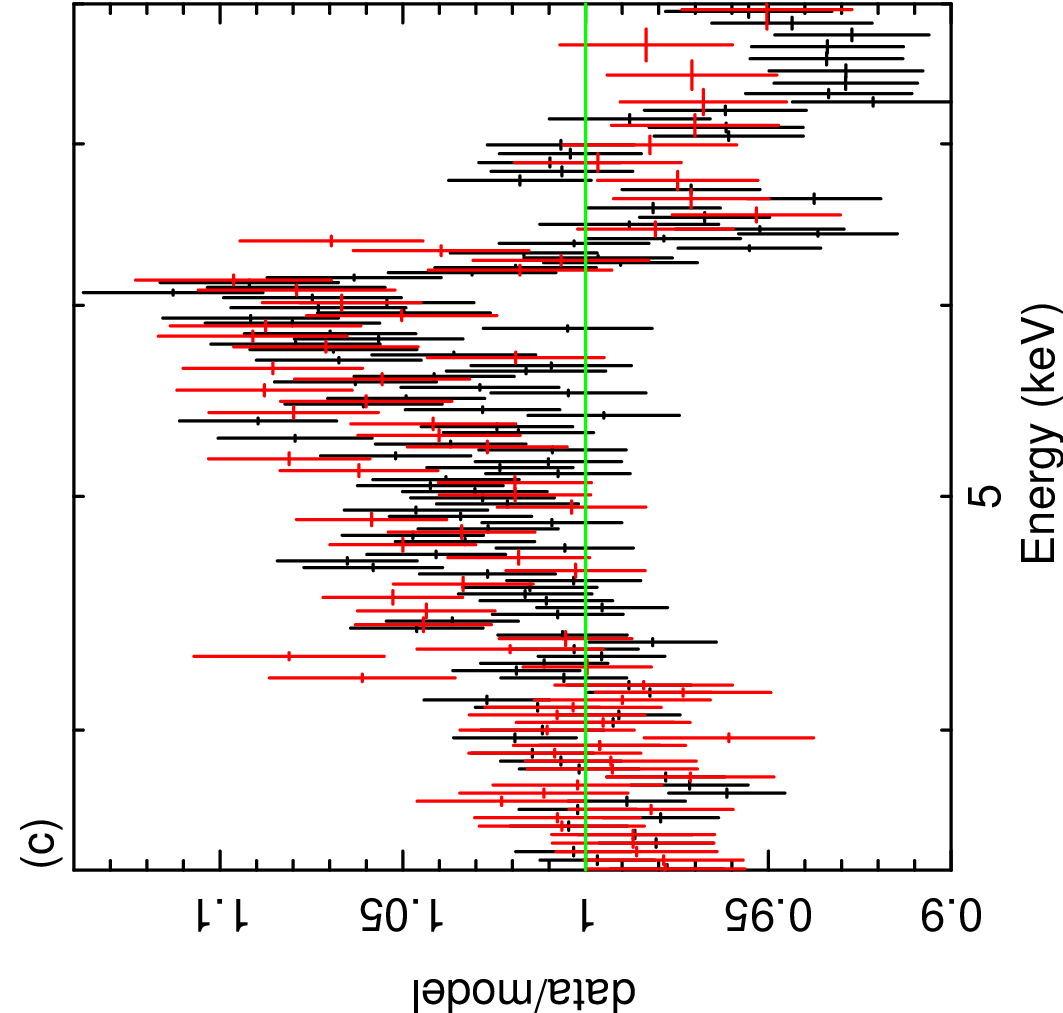}
}
\caption{{\small {\it Left panel}: Simple power-law fit to the
    $3.5-45 \keV$ XIS-FI+PIN spectrum.   {\it Middle panel}: Zoom-in on
    the 4--8\,keV region of the simple power-law fit.   Note the probable
    ``Compton shoulder" on the immediate low-energy side of the 
    strong 6.4\,keV emission line.   {\it Right panel}: Residuals
    remaining when the broad iron line component is removed from a
    simple phenomenological fit to the $3.5-45 \keV$ data (see
    \S\ref{sec:hard}).  In all panels the black points correspond to
    XIS~0+3 data, red to XIS~1 data and blue to PIN data.  The green line
    represents a data-to-model ratio of unity.}}
    \label{fig:hardbandfit}
\end{figure}

It is instructive to begin by examining the hard-band ($3.5-45 \keV$)
XIS+PIN spectrum.  A simple power-law fit to this band reveals
significant spectral complexity (Fig.~\ref{fig:hardbandfit}a,b).  A narrow K$\alpha$
fluorescence line of cold iron ($6.4 \keV$) dominates; structure
redward of this line indicates a possible Compton shoulder as well as
an extended tail reaching down to $\sim 4 \keV$.  Both the narrow
iron line and the broad red wing likely originate from X-ray
reflection and, hence, the convex spectrum between $8-40 \keV$
is readily interpreted as the associated Compton reflection hump.
Structure above the $6.4 \keV$ line indicates a strong absorption
feature at $\sim 6.6 \keV$ and/or an emission line at $\sim 7 \keV$
(likely corresponding to a blend of the K$\beta$ line of cold iron and the
Ly$\alpha$ line of Fe\,{\sc xxvi}).

%
%
Guided by these identifications, we construct an heuristic model of
the hard spectrum consisting of a power-law continuum, a narrow 
  Fe\,{\sc xxvi} emission line (modeled as a Gaussian line centered at
$6.97 \keV$ with $\sigma=10\eV$), reflection from distant,
low-velocity, cold matter (described by the {\tt pexmon}
model\footnote{The {\tt pexmon} model \citep{Nandra2007} is a
  modification of the commonly used {\it pexrav} model
  \citep{Magdziarz1995} which, in addition to the Compton
  backscattered reflection continuum, also models the K$\alpha$ and
  K$\beta$ emission lines of iron, the Compton shoulder of the iron
  K$\alpha$ line, and the K$\alpha$ line of nickel.  The lines are
  included at the appropriate normalization for the assumed
  inclination, abundance, and reflection fraction.  Hence, {\tt
    pexmon} is superior to the usual ``{\tt pexrav}$+${\tt gaussian}"
  model since the strengths of the Compton reflection continuum and
  fluorescent lines are forced to be self-consistent.  Assumptions do
  need to be made, however, when employing this model.  In particular,
  we fix the inclination parameter of this component to be
  $i=60^\circ$ and assume cosmic abundances.}), and a relativistically
broadened cold iron K$\alpha$ line (described by the {\tt laor} model;
\citealt{Laor1991}).   
This model produces an excellent fit to the data ($\chi^2/{\rm
  \nu}=573/544\,(1.05)$) with the following parameters: photon index
$\Gamma=1.68^{+0.01}_{-0.01}$, reflection fraction ${\cal
  R}=0.87^{+0.02}_{-0.06}$, emission line equivalent widths $W_{\rm
  FeXXVI}=28^{+4}_{-5}\eV$, $W_{\rm  broad}=263^{+23}_{-23}\eV$, inner
edge of line emitting disk $r_{\rm in}=3.0^{+0.1}_{-0.8}\,r_g$, index
of line emissivity across disk $q=3.31^{+0.06}_{-0.09}$, and disk
inclination $i<9 \degmark$.  If we replace the {\tt pexmon} model with
a {\tt pexrav} and three separate Gaussian lines for narrow Fe
K$\alpha$ ($6.4 \keV$, $\sigma=0.015 \keV$), Fe K$\beta$ ($7.06 \keV$,
$\sigma=0.015 \keV$) and Compton shoulder ($6.25 \keV$, $\sigma=0.1
\keV$), their equivalent widths are $W_{\rm  K\alpha}=98^{+5}_{-5} \eV$, $W_{\rm
  K\beta} \leq 8 \eV$, and $W_{\rm CS}=22^{+8}_{-6} \eV$, respectively.
We note that, in this fit, the intrinsic widths of the iron lines were taken from
their {\it Chandra}/HETG values \citep{Yaqoob2005}.  Substituting the
individual Gaussian lines and {\tt pexrav} component for the {\tt
  pexmon} model results in a modest change in the
global goodness-of-fit ($\Delta\chi^2/\Delta\nu=-32/-3$), likely owing to
the free normalizations of the emission lines (they are set at fixed
ratios within {\tt pexmon}).  No statistically significant change is
seen in the other model parameters. 

For illustrative purposes only, Fig.~\ref{fig:hardbandfit}c shows the
residuals that remain
when the broad iron line is removed from this spectral model, and the
remaining model parameters are re-fit.  An obvious broad line remains.
However, there are two reasons why this cannot be interpreted as the
``the broad line profile" for this object.  Firstly, NGC~3783 has a
well-known, high-column density warm absorber that, while principally affecting
the soft spectrum, can also introduce subtle spectral curvature up to
$10 \keV$ or more.  Secondly, the broadened iron line is just the tip
of the iceberg; especially when the accretion disk is ionized, the
rest of the reflection spectrum has a sub-dominant but significant
contribution that must be considered.  The statistically unlikely value of the disk
inclination derived from the simple fit ($i<9 \degmark$) is a
signal of these issues.  For these reasons, we are forced into global
modeling of the full $0.7-45 \keV$ spectrum.

\subsection{Guidance from the long {\it Chandra}/HETG observation}
\label{sec:hetg}

It is well known that NGC~3783 possesses a high-column density warm
absorber (WA; e.g., \citealt{Reynolds1997}); this is the greatest
complexity we face when modeling the X-ray spectrum of this source.
For guidance, we turn to the long ($900 \ks$) observation of NGC~3783
with the HETGS on {\it Chandra}.   Extensive analyses of the HETG data
have been published \citep{Kaspi2002,Krongold2003,Netzer2003};
however, to retain consistency and utilize the latest calibrations, we
have retrieved these data from the {\sc tgcat}
database\footnote{http://tgcat.mit.edu/} and have reanalyzed the
1st-order MEG+HEG spectra.   

In more detail, we obtain the {\it Chandra}HETG-data for NGC~3783 from
{\sc tgcat} for each of the OBSIDs corresponding to the 900\,ks
campaign and, coadd together spectra for a given order of a given
grating.  As a result, we obtain four spectral files corresponding to
the time-average $\pm $1st order spectrum from each of the HEG and the
MEG.  These were binned to a minimum of 15 photons per spectral bin in
order to validate the use of $\chi^2$ techniques while still
maintaining spectral resolution.   We then jointly analyzed these
spectra, noticing the 0.5--7\,keV range in the MEG data and the
1--7.5\,keV range in the HEG data.  We permitted the overall
cross-normalization between these four spectra to be free parameters;
in all cases, the best-fitting cross-normalization is within 5\% of unity.   

Fitting these data with a power-law modified by the effects of
Galactic absorption ($N_H=9.91\times 10^{20}\pcmsq$; described using
the {\tt phabs} model of XSPEC) results in a very poor fit with
$\chi^2/\nu = 58962/13112\,(4.50)$.  The residuals suggest a soft excess
component, soft X-ray absorption by a WA, and a prominent fluorescent
iron K$\alpha$ line a 6.4\,keV.  The effect of the WA is modeled using
the XSTAR code \citep{Kallman2001}; for an absorber of a given column
density $N_H$ and ionization parameter $\xi$, XSTAR is used to compute
the absorption imprinted on a power-law X-ray spectrum.  We compute a
grid of XSTAR models, logarithmically sampling a range of column
densities in the range  $N_H:10^{20}-10^{24}\pcmsq$ and a range of
ionization parameters in the range $\xi:1-10^4 \ergcmps$, for use in
spectral fitting.   In the construction of the WA  grids it is assumed
that elemental abundances are fixed to solar 
values\footnote{http://heasarc.nasa.gov/lheasoft/xstar/docs/html/node41.html},
and that the
turbulent velocity of the WA is $200\kmps$.  Dramatic improvements in
the goodness-of-fit are found by the inclusion in the model of three
zones of WA.  To begin with, each WA component is included assuming
that the absorbing gas is at rest with respect to NGC~3783; the
improvement in the fit upon the addition of each WA component was
$\Delta\chi^2=-26316, -3096$ and $-2490$.  The inclusion of a fourth
zone led to a much smaller improvement in the fit and hence was deemed
inappropriate.  The residuals from the 3-WA fit do indicate a soft
excess.  Following \citet{Krongold2003}, we model the soft excess with
a blackbody component (this is intended to be a phenomenological, not
a physical, description of the soft excess; see discussion in
\S\ref{sec:global}) resulting in an improvement in the fit of
$\Delta\chi^2/\Delta\nu=-480/-2$ (i.e., $\chi^2/\nu=26582/13104\,(2.03)$).

While providing a decent fit to the global spectrum, the model thus
described leaves prominent unmodeled emission and absorption lines,
the most prominent of which is the iron fluorescent emission line at
6.4\,keV.  Fitting the iron line with a simple Gaussian model improves
the goodness-of-fit by $\Delta\chi^2/\Delta\nu=-593/-3$, with a line
energy $E=6.398\pm 0.002\keV$ (confirming the identification of cold
iron-K$\alpha$), FWHM$=2000\pm 300\kmps$ and equivalent width
$W_{K\alpha}=88\pm 6\eV$.  However, since we believe that this
component originates from reflection, we shall henceforth model it
using {\tt pexmon}; replacing the simple Gaussian with the {\tt
  pexmon} model convolved with a Gaussian velocity profile (with
FWHM$=1800\pm 300\kmps$) results in a slightly better fit
($\Delta\chi^2=-18$).  At the soft end of the spectrum, the K$\alpha$
emission triplet of O\,{\sc vii} (at 0.574\,keV, 0.569\,keV, 0.561\,keV)
as well as the K$\alpha$ emission line of O\,{\sc viii} (at 0.654\,keV)
are clearly visible.   Modeling these as Gaussian lines at the
redshift of NGC~3783 with common velocity width yields a further
improvement in the goodness of fit ($\Delta\chi^2=-218$) with
best-fitting FWHM$=700\pm 150\kmps$ and equivalent widths
$W_{0.574}=26\pm 6\eV$, $W_{0.569}=14\pm 6\eV$,
$W_{0.561}=47\pm 9\eV$, and $W_{0.654}=23\pm 5 \eV$.  

\begin{figure*}
\begin{center}
\psfig{figure=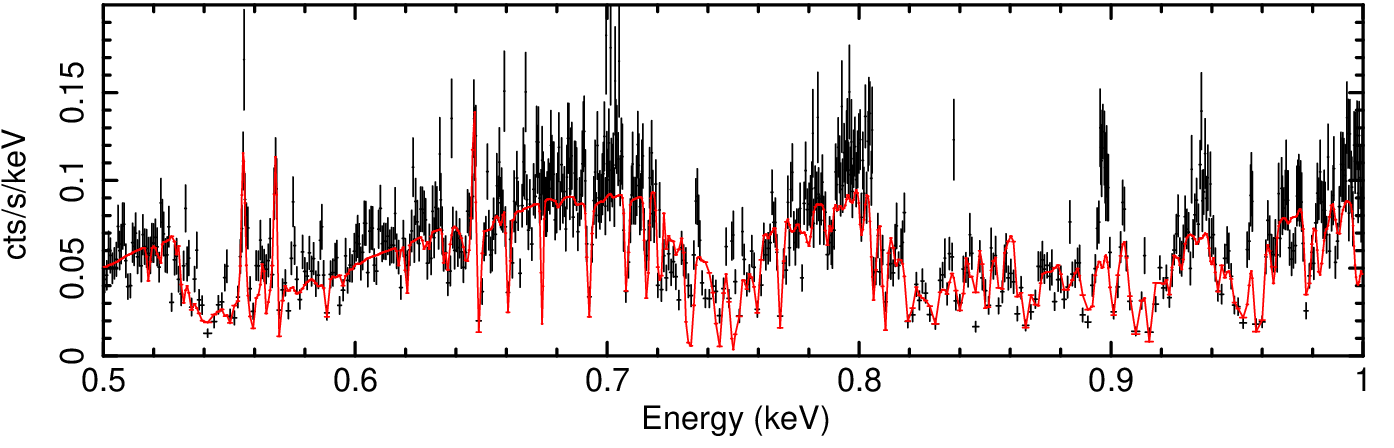,width=0.27\textwidth,angle=270}
\psfig{figure=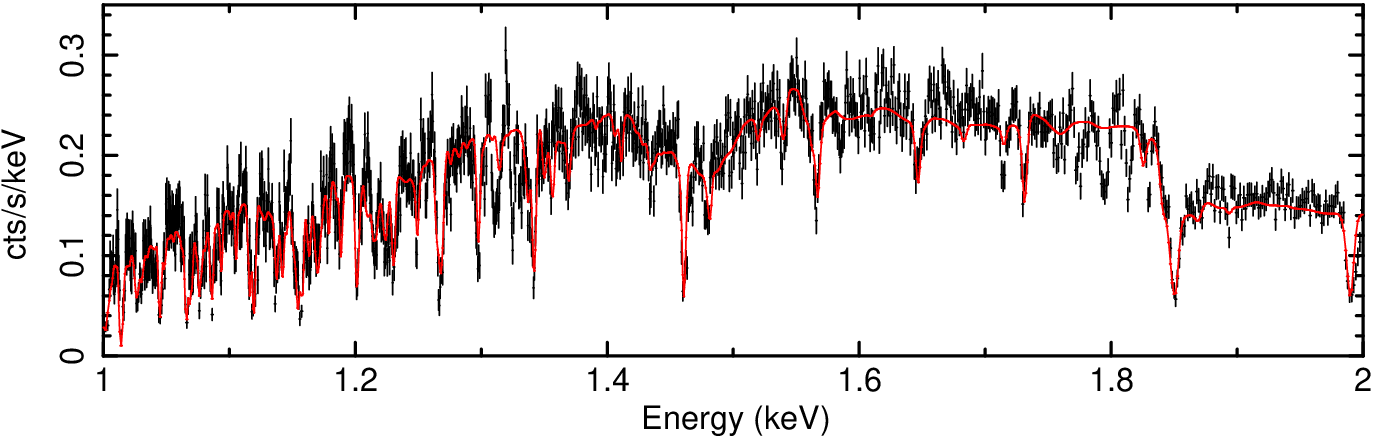,width=0.27\textwidth,angle=270}
\psfig{figure=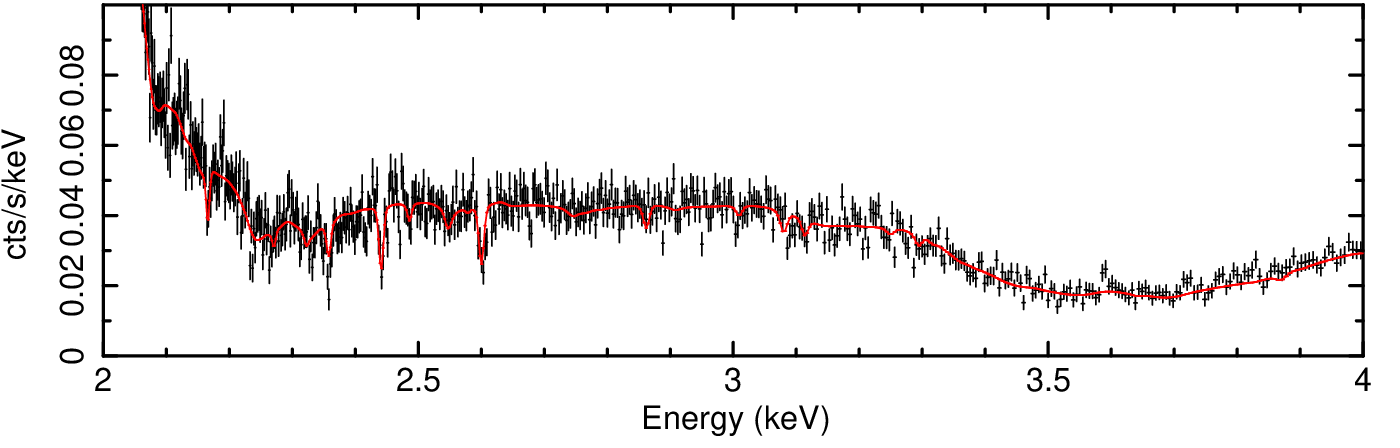,width=0.27\textwidth,angle=270}
\psfig{figure=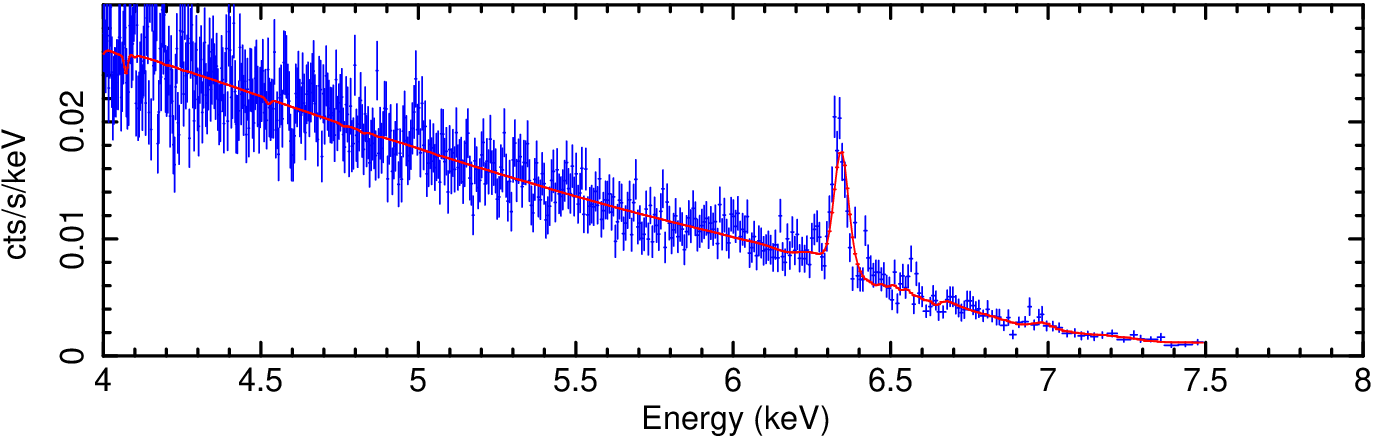,width=0.27\textwidth,angle=270}
\end{center}
\caption{\small{Folded {\it Chandra}/HETG spectrum and best fitting model as
  a function of observed energy.  As described in the text
  (\S\ref{sec:hetg}), the model is fitted simultaneously to the $\pm
  1$ MEG (0.5--7\,keV) and HEG (1--7.5\,keV) data.  However, for
  clarity, we only show here the $-$1-order MEG data (black; first
  three panels) and the $-$1-order HEG data (blue; bottom panel).}}
\label{fig:hetg_fit}
\end{figure*}

It is well known that the WA in this and many other objects
corresponds to outflowing gas.  Relaxing the constraint that the WA
zones are at the systemic redshift of NGC~3783 yields a large
improvement in the fit ($\Delta\chi^2/\Delta\nu=4247/-3$;
$\chi^2/\nu=21532/13094\,(1.64)$), with implied line-of-sight outflow
velocities in the $500-1000\kmps$ range.  These velocities as well as
the other parameters defining the best-fit model for the HETG data are
listed in Table~1.  The spectral model described in this section
(power-law continuum, three-zone WA, blackbody soft excess, reflection
from distant neutral material, and emission lines from {\sc O\,vii}
and {\sc O\,viii}) describes the vast majority of spectral features
seen in the HETG data (see Fig.~\ref{fig:hetg_fit}).

\subsection{Global Modeling of the 0.7--45\,keV Spectrum}
\label{sec:global}

\begin{figure}[t]
\centerline{
\includegraphics[width=0.5\textwidth,angle=270]{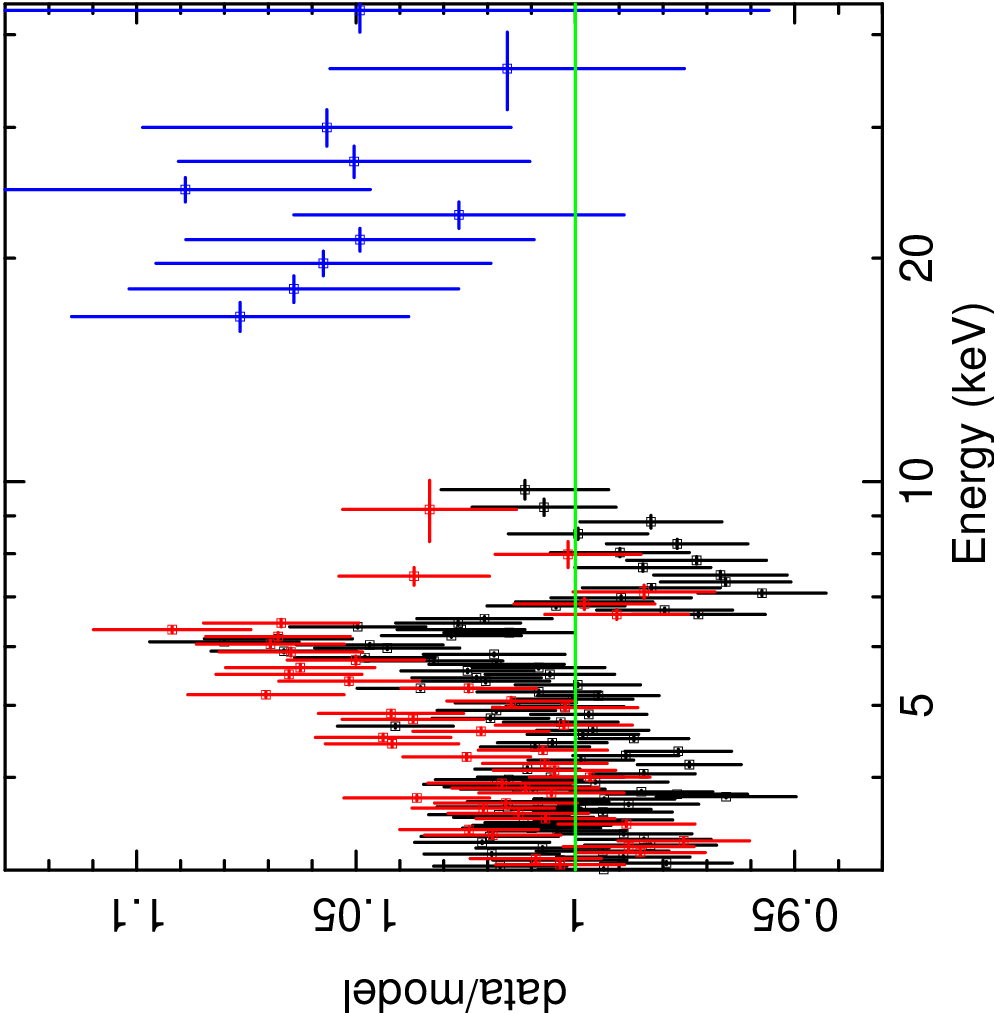}
\includegraphics[width=0.5\textwidth,angle=270]{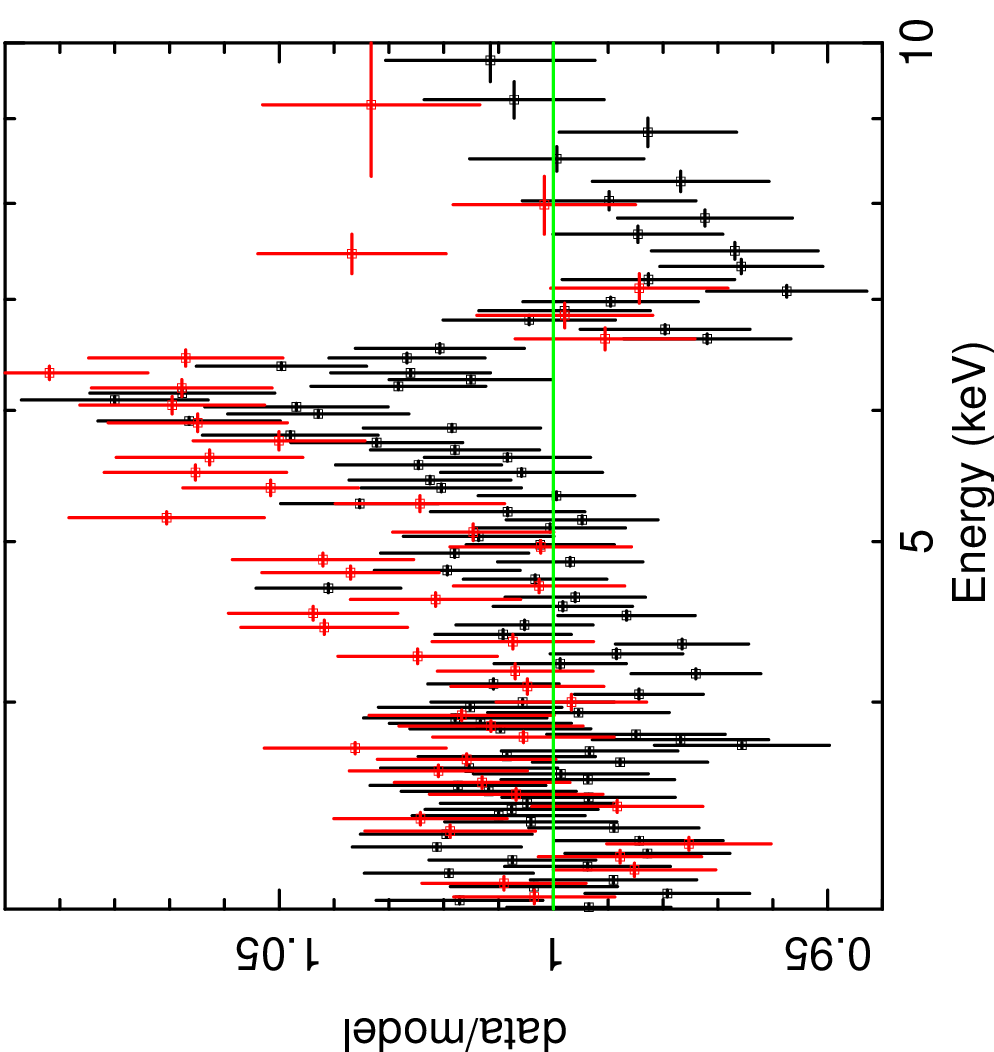}
}
\caption{{\small Results of fitting the XIS+PIN spectrum with a model 
that includes the warm absorbers, distant reflection and scattering/leaked 
soft component but not the relativistic ionized accretion disk.  While
the fitting is performed on the 0.7--45\,keV spectrum, we show for
clarity only the residuals above 3\,keV.   {\it
  Left:} Strong
residuals indicative of a broad iron line and Compton reflection hump are clearly
visible.  This motivates the inclusion of a relativistic disk component into
the spectral model.  The XIS~0+3 data are shown in black and XIS~1
data are in red, while the HXD/PIN data are in blue.  The solid green
line represents a data-to-model ratio of unity.  {\it Right:} Zoom-in
on the Fe K line region.}}
\label{fig:globalmodelingnodisk}
\end{figure}

\begin{figure}[t]
\hbox{
\includegraphics[width=0.5\textwidth,angle=270]{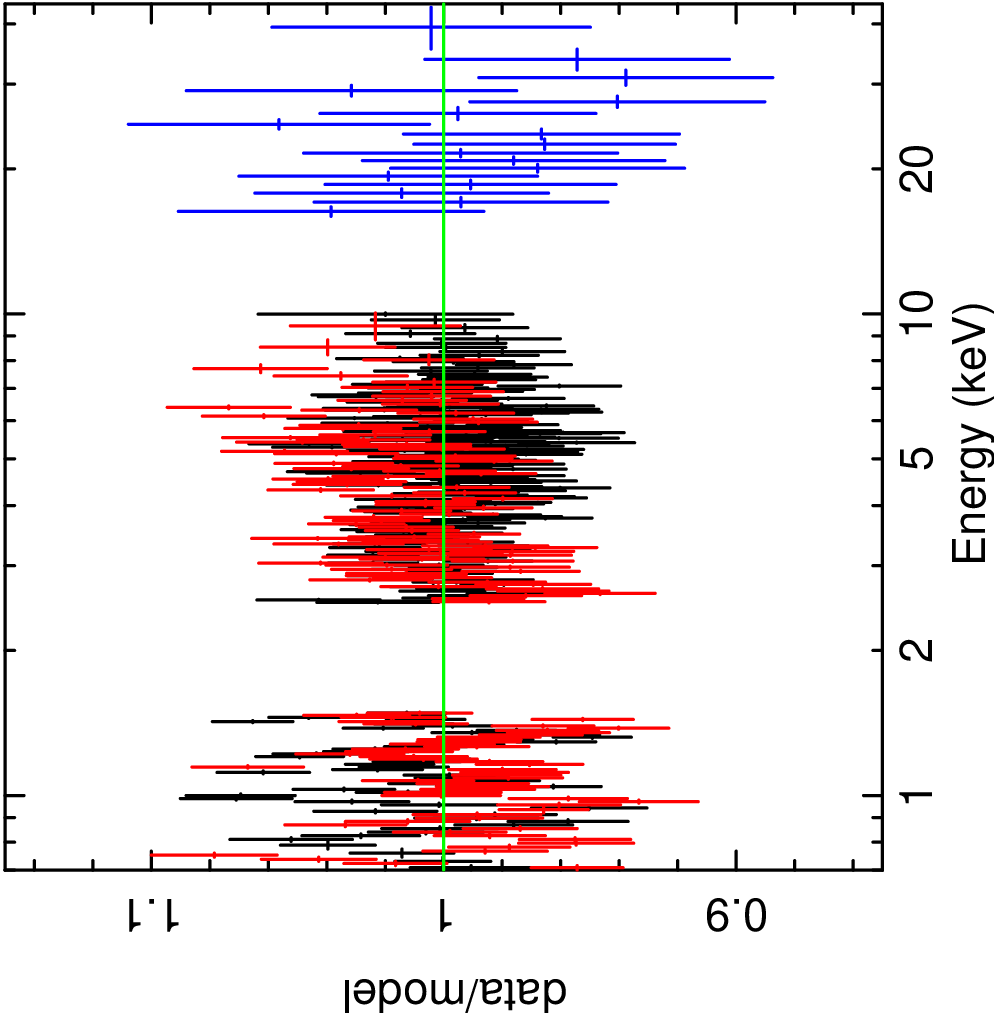}
\includegraphics[width=0.5\textwidth,angle=270]{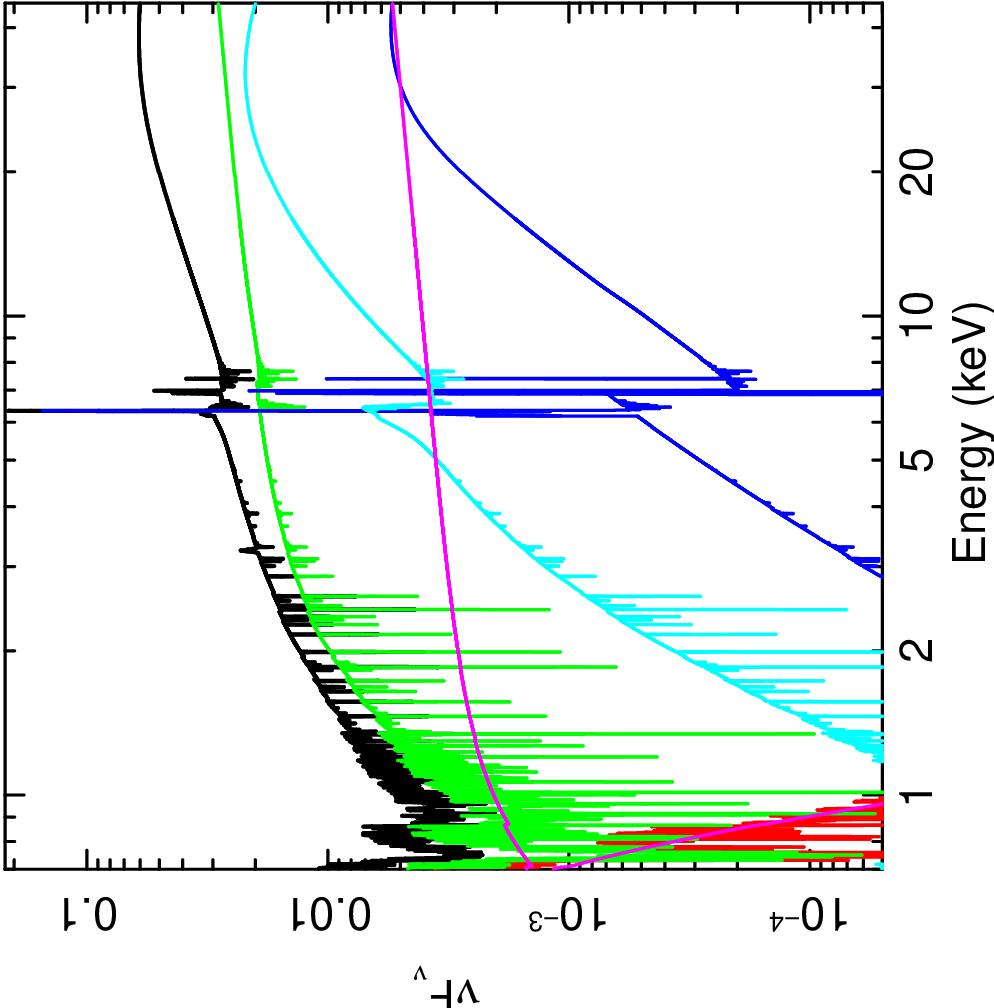}
}
\caption{{\small Global modeling of the $0.7-45 \keV$ XIS-FI+PIN data.
 The {\it left panel} shows the resulting
 residuals from fitting the model (including the relativistic 
 accretion disk) discussed in \S\ref{sec:global}.  Data point colors are as in
 Fig.~2.  The
{\it right panel} shows the best-fitting model 
color coded as follows: (a) green line, 
continuum power-law emission; (b) dark blue line, cold and ionized
iron line emission from distant matter; (c) red line, soft excess
modeled as blackbody; (d) magenta line, significant emission that
scatters around or leaks through the warm absorber; (e) light blue
line, relativistically-smeared disk reflection, (f) thick black
line, total summed model spectrum.  Warm absorption affects all
components apart from (d).}}
\label{fig:globalmodeling}
\end{figure}

To extract the maximal information from the full-band (0.7--45\,keV)
{\it Suzaku} spectrum of NGC~3783, we must compare the data to a
global spectral model which is as physically self-consistent and
realistic as possible.  In constructing this global model, we draw
guidance from our heuristic analysis of the hard-band spectrum
(\S\ref{sec:hard}) as well as the results from the {\it Chandra}/HETG
(\S\ref{sec:hetg}).  The primary continuum emission is taken to be a
power-law (photon index $\Gamma$) with a soft excess which we describe
as a blackbody (temperature $T$).  X-ray reflection of this continuum
from cold, distant material (possibly associated with the
dusty/molecular torus of unified Seyfert schemes) is described using
the {\tt pexmon} model (see \S\ref{sec:hard}).  As discussed in
\S\ref{sec:hard}, the inclination of the {\tt pexmon} is fixed at
$i=60\degmark$ and the abundances are fixed to be solar, as defined in
\citet{Nandra2007}.  These
emission components are then absorbed by a three-zone WA modeled using
the XSTAR tables described in \S\ref{sec:hetg}; the column density
$N_W$ and ionization parameter $\xi$ of each zone is taken to be a
free parameter rather than being fixed to the HETG value.   Since the
{\it Suzaku}/XIS detectors do not have the spectral resolution capable
of constraining the outflow velocities of the various WA zones, we
have elected to hold the redshifts of these components fixed at the
cosmological value for NGC~3783.  Statistically indistinguishable
results are obtained if we, instead, fix the outflow velocities to the
HETG-derived values.  For completeness, we also allow for some
fraction $f_{\rm sc}$ of the continuum to be scattered around (or leak
through) the WAs, i.e., our model allows for ``partial covering". 

Fitting this model to the 0.7--45\,keV {\it Suzaku} data results in a
poor fit ($\chi^2/\nu=1206/679\,(1.78)$) and strong residuals which
indicate the presence of the broad iron line as well as additional
reflection beyond that associated with the narrow iron line
(Fig.~\ref{fig:globalmodelingnodisk}).  This leads us to include
relativistically smeared reflection from an ionized accretion disk
into the spectral model; operationally, we use the ionized reflection
model {\tt reflionx} \citep{Ross2005} convolved with the variable spin
relativistic smearing model {\tt relconv} \citep{Dauser2010}. The {\tt
  relconv} model is a further evolution of the {\tt kerrconv} model of
\citet{Brenneman2006}, employing faster and more accurate
line-integration schemes and allowing black hole spin to be fit as a
free parameter for prograde, non-spinning and retrograde spins ($a \in
[-0.998, 0.998]$).  While the fit achieved with this blurred
reflection model is not statistically ideal
($\chi^2/\nu=917/664\,(1.38)$), there are no broad-band residuals
(Fig.~\ref{fig:globalmodeling}a), and much of the contribution to the
excess $\chi^2$ originates from fine details of the WA-dominated
region below 1.5\,keV.  The model is shown in
Fig.~\ref{fig:globalmodeling}b, and the best-fitting parameter values
are shown in Table~1.

The parameters defining the best-fitting model for the $0.7-45 \keV$
data are shown in Table~\ref{tab:bigtab} along with their 90\%
confidence ranges.  Under the assumption that we can identify the
low-, medium-, and high-ionization components seen in the 2001-HETG
observation with those seen in out 2009-{\it Suzaku} data, we see that
both the column density and the ionization state of the low-ionization
absorber have increased somewhat ($\Delta\log\xi\approx 0.28$, $\Delta
N_{\rm WA}\approx 3\times 10^{21}\pcmsq$).  In contrast, the medium-
and high-ionization absorbers have slightly dropped in ionization
parameter.  Given that these different zones are likely at very
different distances from the central engine with very different plasma
densities, they will possess very different
recombination/photoionization timescales and hence will respond to
changes in the ionization flux on different timescales.  Thus, it is
not surprising that we see a mixture of increasing and decreasing
ionization states in the various WA zones.

We also note a change in the both the temperature and normalization of
the blackbody component between the 2001-HETG and 2009-{\it Suzaku}
data.  This merits some discussion.  The blackbody component used to
phenomenologically parameterize the soft excess was first employed by
\citet{Krongold2003}, who found $kT=0.10 \pm 0.03 \keV$ and $A_{\rm
  bb}=2.0 \pm 0.7 \times 10^{-4}$, where the normalization is in units
of $L_{39}/D^2_{10}$ ($L_{39}$ is luminosity of the component in units
of $10^{39} \ergps$ and $D_{10}$ is distance to the source in units of
$10 \kpc$).  Our analysis of the same HETG data confirms the
\citet{Krongold2003} result.  By contrast, our analysis of the {\it
  Suzaku}/XIS+PIN spectra finds a lower temperature
($kT=0.060^{+3}_{-4}\keV$) and a normalization that is almost two
orders of magnitude greater ($A_{\rm bb}=8.4^{+6.0}_{-2.7}\times
10^{-3}$).  We stress that neither the use of a blackbody to model the
soft excess nor the precise change in the parameters of the blackbody
should be interpreted literally.  In particular, the significant change
in the normalization of this component is misleading --- the lower
energy cutoffs in both the HETG analysis (0.5\,keV) and the XIS
analysis (0.7\,keV) are much higher than the peak of this blackbody
component and, thus, only the Wien tail of this component is playing
any role in the spectral fitting.  Given this fact, even a modest drop
in the temperature must
be compensated for by a large increase in normalization in order to
have a comparable contribution in the observed energy band.  While the
physical nature of the soft excess is of intrinsic interest, it is
beyond the scope of this paper.  We have verified that different
treatments of the soft excess (replacing the blackbody spectrum with
bremsstrahlung or a steep power-law component) do not affect the
interpretation of the spectrum above 2\,keV. 

The principal focus of this work is the signature of the
relativistic accretion disk.  Our global fit finds reflection from a
rather low-ionization accretion disk ($\xi<9\ergcmps$) extending down
to the innermost stable circular orbit (ISCO) of a rapidly rotating
black hole ($a \geq 0.98$).  The
emissivity/irradiation profile defining the reflection spectrum,
modeled as a broken power-law, is found to have an inner power-law
index of $q_1=5.2^{+0.7}_{-0.8}$ breaking to $q_2=2.9\pm 0.2$ at a
radius of $r_{\rm br}=5.4^{+1.9}_{-0.9}r_g$.  If these indices are
tied together in the model, i.e., if $q_1=q_2$, the fit worsens
considerably ($\Delta\chi^2/\Delta\nu=+21/+1$, with $q_1=q_2=3.0 \pm
0.3$) and the black hole spin is also less tightly constrained: $a
\geq 0.25$.  

The iron abundance of
the disk has been constrained to lie between $2.8-4.6$ times solar.
To probe the robustness of this constraint we have refitted the data
in three different ways, each allowing for slight differences in the
way the iron abundance was handled in our model: (1) fixing Fe/solar
of the distant reflector ({\tt pexmon}) to that of the inner disk
reflection ({\tt reflionx}), with both values frozen at Fe/solar$=1$;
(2) allowing these linked abundances allowed to vary freely; and (3)
allowing both abundances to vary freely and independently.  Compared
with the global best-fit, scenario (1) resulted in a worsening of
goodness-of-fit by $\Delta\chi^2=33$ and unconstrained black hole spin
at the 90\% confidence level, scenario (2) resulted in a
marginal decrease in goodness-of-fit by $\Delta\chi^2=7$ (no change in
spin constraints) and scenario
(3) yielded no change in the goodness-of-fit (no change in spin
constraints).  In summary, the high iron
abundance of the {\tt reflionx} is statistically preferred to the
solar value; the high spin value is dependent upon the high iron
abundance, but the high abundance is strongly preferred in the fit.  
Because the iron abundance of the distant reflector 
could not be constrained independently of the relativistic reflector,
the {\tt pexmon} and {\tt reflionx} iron abundances have been linked
in our best model fit.

\begin{figure}
\hbox{
\psfig{figure=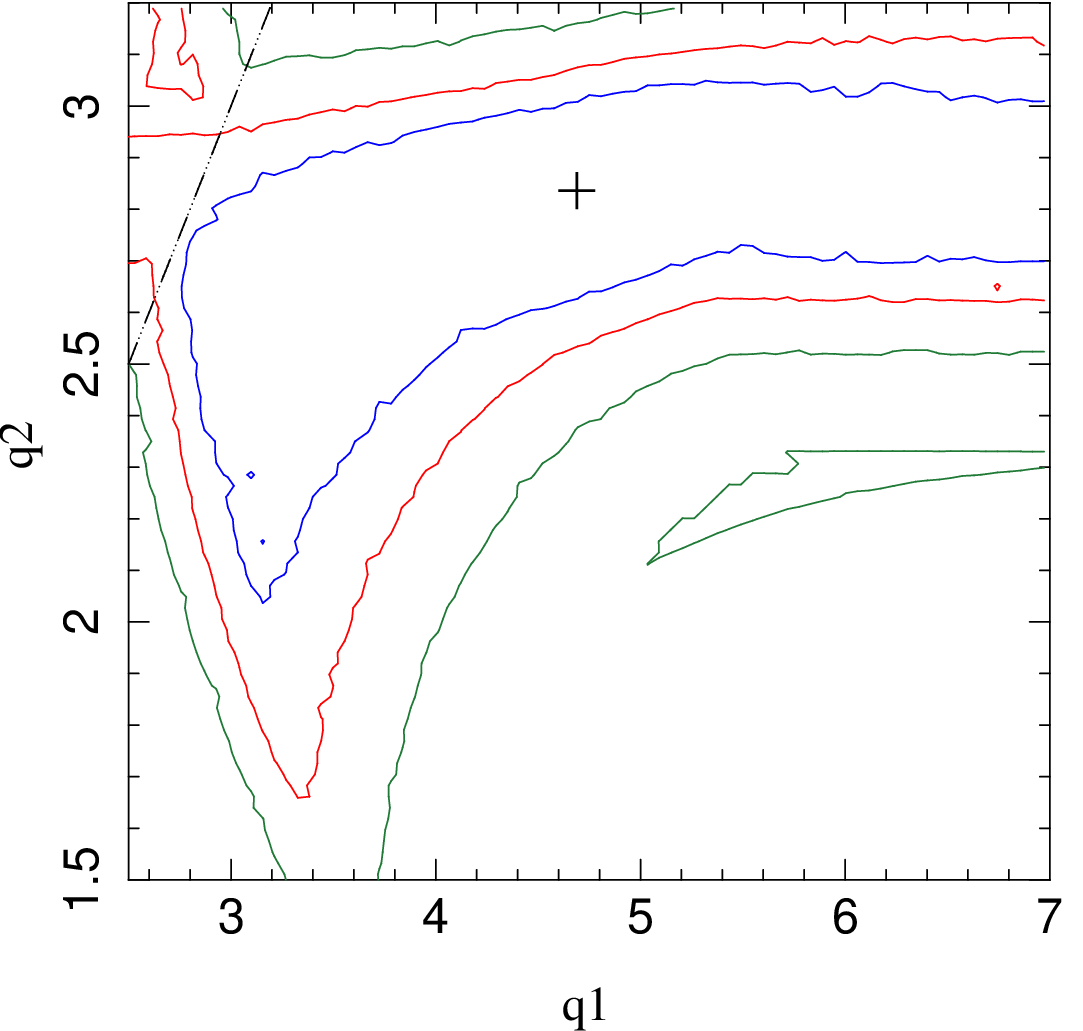,width=0.45\textwidth}
\psfig{figure=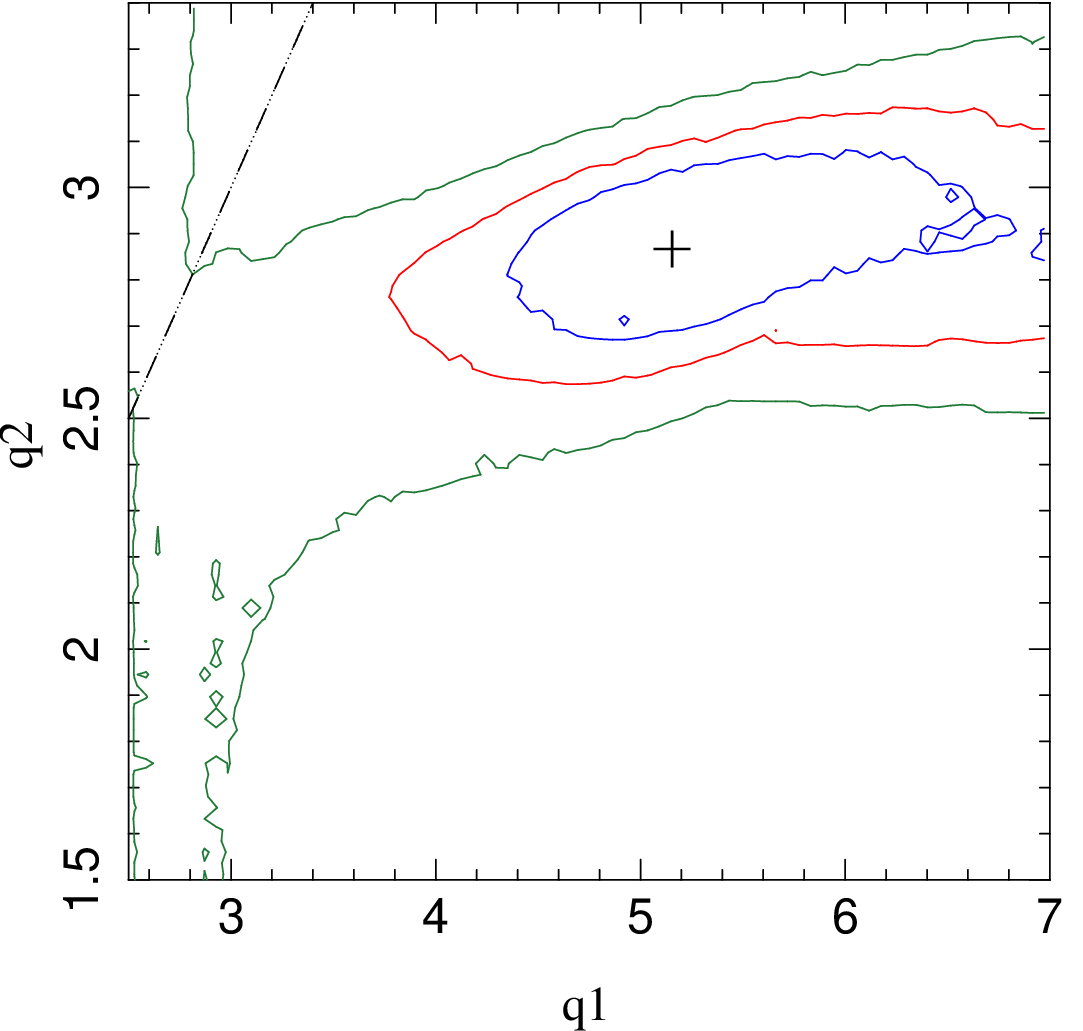,width=0.45\textwidth}
}
\caption{\small{Confidence contours on the $(q_1,q_2)$-plane for the
  3--45\,keV band fit assuming a XIS/PIN cross-normalization that is
  free (left panel) or fixed at the nominal value of 1.18 (right
  panel).   Blue, red and green lines show the 68\%, 90\% and 99\% confidence
  contours for two interesting parameters
  ($\Delta\chi^2=2.3,\,4.6,\,9.2$), respectively.  The dot-dashed
  black line represents $q_1=q_2$.}}
\label{fig:qcontours}
\end{figure}

To gauge the importance of the (complex) soft spectrum on our global
fit, we have also conducted a restricted hard-band (3--45\,keV) fit.
Since a hard-band fit cannot constrain the parameters of the WA or
soft excess, these parameters are constrained to lie within their 90\%
confidence ranges as derived from the 0.7--45\,keV analysis.  To be
most conservative, we also relax the constraint on the XIS/PIN
cross-normalization, allowing it to be a free parameter.  The
resulting fit is listed in the last column of Table~\ref{tab:bigtab}.
For this fit $\chi^2/\nu=499/527\,(0.95)$, a great improvement over
the $0.7-45 \keV$ fit, and confirmation that the small residuals below
$\sim 1.5 \keV$ are the primary contribution to the large reduced
$\chi^2$ of the full spectral fit. While the parameter values are
equivalent to those of the $0.7-45 \keV$ fit within errors, the
uncertainties on the parameters are larger when only the hard spectrum
is considered.  This is especially true for the inner disk emissivity
and break radius of the {\tt relconv} model, which exhibit a strong
degeneracy without the soft spectrum data.  Figure~\ref{fig:qcontours}
shows the confidence contours on the $(q_1,q_2)$-plane for this
hard band fit; we see that the outer disk emissivity index, $q_2$, is
well-constrained, whereas
the constraints on the inner disk emissivity index, $q_1$, are clearly
worse.  Fixing the XIS/PIN
cross-normalization at the nominal value of 1.18 tightens the
constraints but still leaves a significant degeneracy between $q_1$
and $r_{\rm br}$.  
(Fig.~\ref{fig:qcontours}).  

Using the best-fitting $0.7-45 \keV$ spectral model with the XIS~0 normalization, the $2-10 \keV$
observed-frame flux of NGC~3783 is $F_{2-10}=6.04\times 10^{-11}\ergpcmsqps$.  Adopting a
standard cosmological model ($H_0=71\kmpspMpc$, $\Omega_M=0.3$,
$\Omega_\Lambda=0.7$), this implies a rest-frame luminosity of $L_{2-10}=1.26\times
10^{43}\ergps$.  The hard X-ray band yields a $16-45 \keV$ flux of
$F_{16-45}=1.07 \times 10^{-10}\ergpcmsqps$ for a rest-frame luminosity of
$L_{16-45}=2.24 \times 10^{43}\ergps$. 

{\small
\begin{table}
\hspace{-1.0cm}
\begin{tabular}{|lllll|}\hline\hline
{\bf Model component} & {\bf Parameter} & {\bf HETG} & {\bf Suzaku} ($0.7-45 \keV$) & {\bf Suzaku} ($>3 \keV$) \\\hline
Galactic column & $N_{\rm H}$ & $9.91(f)$ & $9.91(f)$ & $9.91(f)$ \\\hline
WAbs1 & $N_{\rm WA}$ & $51.7^{+0.8}_{-0.7}$ & $90^{+10}_{-14}$ & $90(f)$\\
      &	$\log\,\xi$ & $1.15^{+0.01}_{-0.01}$ & $1.47^{+0.03}_{-0.03}$ & $1.47(f)$\\
      & $\Delta z$ & $-(1.4^{+0.07}_{-0.07})\times 10^{-3}$ & $0(f)$ & $0(f)$\\\hline
WAbs2 & $N_{\rm WA}$ & $127^{+1.8}_{-2.0}$ & $159^{+31}_{-21}$ & $159(f)$\\
      & $\log\,\xi$ & $2.08^{+0.01}_{-0.01}$ & $1.93^{+0.02}_{-0.01}$ & $1.93(f)$\\
      &	$\Delta z$ & $-(1.0^{+0.3}_{-0.3})\times 10^{-3}$ & $0(f)$ & $0(f)$\\\hline
WAbs3 & $N_{\rm WA}$ & $268^{+10}_{-12}$ & $168^{+48}_{-42}$ & $168(f)$\\
      &	$\log\,\xi$ & $2.83^{+0.01}_{-0.01}$ & $2.53^{+0.05}_{-0.02}$ & $2.53(f)$\\
      &	$\Delta z$ & $-(3.4^{+0.4}_{-0.4})\times 10^{-4}$ & $0(f)$ & $0(f)$\\\hline
PL    &	$\Gamma$ & $1.62^{+0.01}_{-0.01}$ & $1.81^{+0.10}_{-0.05}$ & $1.84^{+0.06}_{-0.05}$\\
      &	$A_{\rm pl}$ & $(1.49^{+0.01}_{-0.01})\times 10^{-2}$ & $(1.46^{+0.09}_{-0.04}) \times 10^{-2}$ & $(1.52^{+0.10}_{-0.08}) \times 10^{-2}$ \\\hline
BB    &	$kT$(eV) & $107^{+3}_{-3}$ & $60^{+3}_{-4}$ & $60(f)$\\
      &	$A_{\rm bb}$ & $(1.4^{+0.1}_{-0.1})\times 10^{-4}$ & $(8.45^{+5.98}_{-2.67})\times 10^{-3}$ & $8.45 \times 10^{-3}(f)$ \\\hline
Scattered fraction & $f_{sc}$ & $(2.3^{+0.4}_{-1.0})\times 10^{-2}$ & $0.17^{+0.02}_{-0.02}$ & $0.17(f)$\\\hline
Cold Reflection & $\cal R_{\rm cold}$ & $0.49^{+0.04}_{-0.03}$ & $0.46^{+0.12}_{-0.07}$ & $0.62^{+0.31}_{-0.20}$ \\
                & PL cutoff (keV) & -- & $200(f)$ & $200(f)$ \\\hline
GAU line & $E$ (keV) & -- & $6.97(f)$ & $6.97(f)$\\
	 & $\sigma$ (keV) & -- & $0.0154(f)$ & $0.0154(f)$\\
	 & $W_{\rm FeXXVI}$ (eV) & -- & $22^{+5}_{-5}$ & $18^{+6}_{-5}$\\
         & $A_{\rm FeXXVI}$ & -- & $(1.28^{+0.29}_{-0.31}) \times 10^{-5}$ & $(1.11^{+0.39}_{-0.31}) \times 10^{-5}$ \\\hline 
Accretion disk & $Z_{\rm Fe}$ & -- & $3.7^{+0.9}_{-0.9}$ & $2.2^{+2.4}_{-0.9}$\\
	& $\xi$	& -- & $<8$ & $<67$\\
        & $\cal R_{\rm rel}$ & -- & $0.21^{+1.56}_{-0.07}$ & $0.23^{+14.66}_{-0.03}$ \\\hline
	& $i$ & -- & $22^{+3}_{-8}$ & $19^{+6}_{-14}$\\
	& $r_{\rm in}$ & -- & ISCO(f) & ISCO(f)\\
	& $q_1$ & -- & $5.2^{+0.7}_{-0.8}$ & $4.7^{+1.9}_{-1.2}$\\
	& $r_{\rm br}$ & -- & $5.4^{+1.9}_{-0.9}$ & $6.0^{+16.9}_{-1.9}$\\
	& $q_2$ & -- & $2.9^{+0.2}_{-0.2}$ & $2.8^{+0.3}_{-0.5}$\\
	& $r_{\rm out}$	& -- & $400(f)$	& $400(f)$\\\hline
PIN/XIS norm &	& -- & $1.18(f)$ & $1.15^{+0.07}_{-0.07}$\\\hline
SMBH spin & $a$	& -- & $\geq 0.98$ & $0.98^{+0.02}_{-0.34}$\\\hline
$\chi^2/\nu$ & & $21532/13094\,(1.64)$ & $917/664\,(1.38)$ & $499/527\,(0.95)$ \\\hline\hline
\end{tabular}
\hspace{-1.0cm}
\caption{\small{Spectral fit parameters.  All errors are quoted at the 90\%
  confidence level for one interesting parameter
  ($\Delta\chi^2=2.7$).  Parameters marked with an ``(f)'' had their
  values fixed during the fit.
 Units of normalization are in $\phpcmsqps$, column density is in
  units of $10^{20} \pcmsq$, ionization parameter is in $\ergcmps$, iron abundance is
  relative to solar (linked between the {\tt reflionx} and {\tt
  pexmon} reflection components), inclination is in degrees, radii are
  in $r_{\rm g}$, and spin parameter is
  dimensionless, but is defined as $a \equiv cJ/GM^2$.  See \S3 for
  details.}}
\label{tab:bigtab}
\end{table}
}

\subsection{The Spin of the Black Hole}
\label{sec:spin}

\begin{figure}[t]
\centerline{
\includegraphics[width=0.6\textwidth,angle=270]{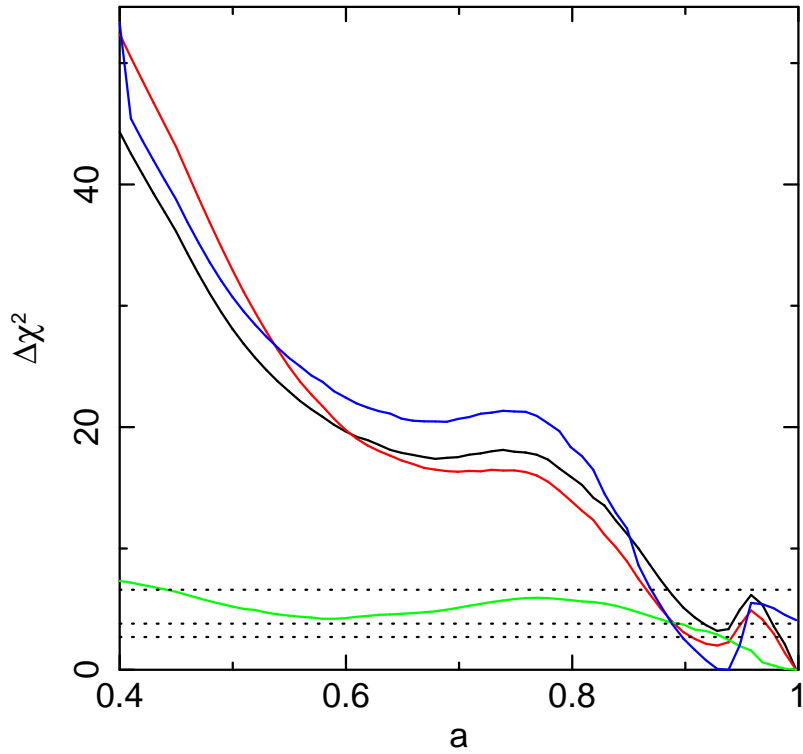}
}
\caption{{\small Goodness-of-fit parameter $\Delta\chi^2$ as a
    function of the assumed black hole spin for our fiducial model (black 
    line), the fiducial model except with frozen WA parameters (red line), 
    the fiducial model except with free XIS/PIN normalization (blue line),
    and a hard-band ($>3$\,keV) fit only (green line).  Confidence
    levels are indicated with horizontal black lines and are derived
    for one interesting parameter. 
        See text in \S3.4 for details.}}
\label{fig:spinconstraints}
\end{figure}

Our fiducial spectral model discussed above yields a spin constraint
of $a \geq 0.98$ ($90\%$ confidence), or $a \geq 0.88$ ($99\%$ confidence).
However, given the subtle nature of the spin measurements, it is
useful to address the systematic issues that may be introduced by the
modeling and analysis techniques. 

We can assess the role of different analysis-related assumptions on
the derived spin by comparing the variation of $\chi^2$ with $a$.
Figure~\ref{fig:spinconstraints} (black line) shows 
$\Delta\chi^2(a)=\chi^2(a)-\chi^2_{\rm best-fit}$ from our fiducial 
analysis that underlies the constraint just quoted. It is interesting
to note the non-monotonic nature of the $\chi^2$-space above
$a\sim 0.75$.  We consider a few variants from this fiducial analysis
in order to probe the sensitivity of the spin measurement.  An
important issue is the extent to which the warm absorber parameters
are trading-off with the derived black hole spin.   Thus, we repeat
the analysis with the warm absorber parameters (column densities
and ionization parameters for all three zones) fixed at their best-fit
values from the fiducial model.  The resulting spin constraints
are shown in Fig.~\ref{fig:spinconstraints} ($a \geq 0.98$; red line) and are very 
similar to the fiducial model, indicating little or no degeneracy between
spin and the warm absorber parameters.   

Secondly, to the extent that the strength of the Compton reflection hump
is important, we may be concerned about the effect of cross-calibration
errors in the flux normalization between the XIS and the PIN spectra.  Thus,
we repeat the spectral analysis, leaving the cross-normalization factor as
a free parameter.  The best-fit value is slightly smaller than our fiducial
value ($1.15$ vs. $1.18$), but the improvement in the goodness-of-fit is
only marginally significant ($\Delta\chi^2=6$ for one additional degree
of freedom).  The spin is constrained to be slightly smaller than that
of the fiducial model (to $90\%$ confidence, $a=0.92-0.95$;
Fig.~\ref{fig:spinconstraints}, blue line).  

Lastly, we may be concerned that the spin fits are being
driven by the contribution of the ionized disk to the high-S/N by
highly-complex region of the spectrum below $1.5 \keV$.  Thus, we have
repeated our analysis including data only above $3 \keV$.  Here, too, we
allow the XIS/PIN cross-normalization to be a free parameter.  Given the
lack of data at soft energies to constrain them, the WA, BB and
scattered fraction components were frozen to
their best-fitting values for the full-band, free cross-normalization case 
(which is, within errors, identical to the WA parameters for the fiducial 
model).   Yet again, the best-fit spin parameter is similar to
that of the fiducial
model ($a \geq 0.95$; Fig.~\ref{fig:spinconstraints}, green line).
This indicates that the fitted spin
value is indeed driven by the Fe K band.

\section{Discussion and Conclusions}

The X-ray spectrum of NGC~3783 is complicated; in addition to the
effects of a multi-zone warm absorber, there are suggestions that some
fraction (17\%) of the primary X-ray emission can scatter around
or leak through the warm absorber.   However, despite this complexity,
the high-S/N and broad bandpass of {\it Suzaku} allows us to robustly
detect and study the relativistically-smeared X-ray reflection
spectrum from the surface of the inner accretion disk.   
Assuming that
the region within the general relativistic radius of marginal
stability does not contribute to the reflection spectrum
\citep{ReynoldsFabian2008} we determine a lower limit of $a \geq 0.98$ (90\% 
confidence) to the dimensionless spin parameter of the black hole.
Even at the 99\% confidence level, we can constrain the spin to be
$a \geq 0.88$.  Relaxing the assumed XIS/PIN cross-normalization
or neglecting the soft-band data (but then freezing the WA 
parameters) allows the model to find a slightly better fit and
 makes the constraints slightly lower ($a=0.92-0.95$, $a \geq 0.95$ at
 90\% confidence, respectively; 
 $a \geq 0.88$, $a\geq 0.90$ at 99\% confidence, respectively). 

{\small
\begin{sidewaystable} 
\renewcommand{\thefootnote}{\emph\alph{footnote}}
\begin{tabular}{|cccccccccc|}\hline\hline
{\bf AGN} & {\bf a} & {\bf W$_{K\alpha}$} & {\bf $q_1$} & {\bf Fe/solar} & {\bf
  $\xi$} & {\bf log M} & {\bf $L_{\rm bol}/L_{\rm Edd}$} & {\bf Host}
  & {\bf WA} \\
\hline
MCG--6-30-15\footnote{Brenneman \& Reynolds (2006), Miniutti \etal (2007).} & $\geq 0.98$ &
  $305^{+20}_{-20}$ & $4.4^{+0.5}_{-0.8}$ & $1.9^{+1.4}_{-0.5}$ &
  $68^{+31}_{-31}$ & $6.65^{+0.17}_{-0.17}$ & $0.40^{+0.13}_{-0.13}$ & E/S0 & yes\\
\hline
Fairall~9\footnote{Schmoll \etal (2009), though note some discrepancies
  with Patrick \etal (2010).} & $0.65^{+0.05}_{-0.05}$ & $130^{+10}_{-10}$ &
  $5.0^{+0.0}_{-0.1}$ & $0.8^{+0.2}_{-0.1}$ & $3.7^{+0.1}_{-0.1}$ &
  $8.41^{+0.11}_{-0.11}$ & $0.05^{+0.01}_{-0.01}$ & Sc & no\\ 
\hline
SWIFT J2127.4+5654\footnote{Miniutti \etal (2009), though note some discrepancies
  with Patrick \etal (2010).} & $0.6^{+0.2}_{-0.2}$ &
  $220^{+50}_{-50}$ & $5.3^{+1.7}_{-1.4}$ & $1.5^{+0.3}_{-0.3}$ &
  $40^{+70}_{-35}$ & $7.18^{+0.07}_{-0.07}$ & $0.18^{+0.03}_{-0.03}$ & --- & yes\\ 
\hline
1H0707--495\footnote{Zoghbi \etal (2010), de La Calle P{\'e}rez \etal (2010).} & $\geq 0.98$ & $1775^{+511}_{-594}$
  & $6.6^{+1.9}_{-1.9}$ & $\geq 7$ & $50^{+40}_{-40}$ &
  $6.70^{+0.40}_{-0.40}$ & $\sim 1.0_{-0.6}$ & --- &no \\
\hline
Mrk~79\footnote{Gallo \etal (2005, 2010).} & $0.7^{+0.1}_{-0.1}$ & $377^+{47}_{-34}$ &
  $3.3^{+0.2}_{-0.1}$ & $1.2*$ & $177^{+6}_{-6}$ & $7.72^{+0.14}_{-0.14}$ & $0.05^{+0.01}_{-0.01}$ & SBb & yes\\
\hline
Mrk~335\footnote{Patrick \etal (2010).} & $0.70^{+0.12}_{-0.01}$ &
  $146^{+39}_{-39}$ & $6.6^{+2.0}_{-1.0}$ & $1.0^{+0.1}_{-0.1}$ & $207^{+5}_{-5}$ &
  $7.15^{+0.13}_{-0.13}$ & $0.25^{+0.07}_{-0.07}$ & S0a & no\\
\hline
NGC~7469\footnotemark[6] & $0.69^{+0.09}_{-0.09}$ & $91^{+9}_{-8}$
  & $\geq 3.0$ & $\leq 0.4$ & $\leq 24$ & $7.09^{+0.06}_{-0.06}$ & $1.12^{+0.13}_{-0.13}$ & SAB(rs)a & no\\
\hline
NGC~3783\footnote{This work.} & $\geq 0.98$ & $263^{+23}_{-23}$ & $5.2^{+0.7}_{-0.8}$ & $3.7^{+0.9}_{-0.9}$ &
  $\leq 8$ & $7.47^{+0.08}_{-0.08}$ & $0.06^{+0.01}_{-0.01}$ & SB(r)ab & yes\\
\hline\hline 
\end{tabular}
\caption{\small{Summary of black hole spin measurements derived from
  relativistic reflection fitting of SMBH spectra.  Data are taken
  with {\it Suzaku} except for 1H0707--495, which was observed with
  {\it XMM-Newton}, and MCG--6-30-15, in which the data from {\it XMM}
  and {\it Suzaku} are consistent with each other.  Spin ($a$) is
  dimensionless, as defined previously.  $W_{K\alpha}$ denotes the equivalent
  width of the broad iron line relative to the continuum in units of
  eV.  Parameter $q_1$ represents the inner disk emissivity index and
  is unitless.  Fe/solar is the iron abundance of the inner disk in
  solar units, while $\xi$ is its ionization parameter in units of
  $\ergcmps$.  $M$ is the mass of the black hole in solar masses, and $L_{\rm
  bol}/L_{\rm Edd}$ is the Eddington ratio of its luminous output.
  Host denotes the galaxy host type and WA denotes the
  presence/absence of a warm absorber.  Values marked with an asterisk
  either were fixed in the fit or have
  unknown errors.  
  All masses are from \citet{Peterson2004} except MCG--6-30-15,
  1H0707--495 and SWIFT J2127.4+5654, which are taken from
  \citet{McHardy2005}, \citet{Zoghbi2010} and \citet{Malizia2008}, respectively.  All bolometric
  luminosities are from \citet{Woo2002} except for the same three
  sources.  The same references for MCG--6-30-15 and SWIFT
  J2127.4+5654 are used, but host types for 1H0707--495 and SWIFT
  J2127.4+5654 are unknown.}}
\label{tab:table2}
\end{sidewaystable}
}

Including this result, four out of eight of the AGN with reliable spin
measurements may have spins greater than $a=0.8$ (see Table~2).  Spin measurements for
more sources are required before we can draw any conclusions about the
spin distribution function, but here we note that there are
potentially important selection effects biasing any flux-limited
sample towards high spin values.  For standard accretion models, the
efficiency of black hole accretion increases as the spin of the black
hole increases.  So, all else being equal, an accreting, rapidly
spinning black hole will be more luminous than an accreting, slowing
spinning black hole and hence will be over-represented in flux-limited
samples.

We illustrate this effect by calculating the selection bias given some
very simple assumptions.   Suppose that a flux limited sample is
constructed in some band B.  The accretion luminosity in that band
will be given by
\begin{equation}
L=K_B \eta \dot{M}c^2,
\end{equation}
where $K_B$ is the fraction of luminosity appearing in band B
(i.e. the reciprocal of the bolometric correction), $\eta$ is the
accretion efficiency, and $\dot{M}$ is the mass accretion rate.  Now
let us assume that $\dot{M}$ has no explicit spin dependence
(e.g. is determined by the larger circumnuclear environment), and that
the spectral energy distribution and hence $K_B$ is independent of
spin.   Thus, the space density of sources with accretion rates in the
range $\dot{M}\rightarrow \dot{M}+d\dot{M}$ and spins in the range
$a\rightarrow a+da$, denoted $\Phi(\dot{M},a)\,d\dot{M}\,da$, can be
taken as a given function set by the astrophysics of black hole growth.

We assume a Euclidean universe, valid for the local/bright AGN
samples relevant for spin measurements with the current generation of
X-ray observatories.  The number of sources in a flux-limited sample
with luminosity in the range $L\rightarrow L+dL$ and spins in the
range $a\rightarrow a+da$ is then,
\begin{equation}
dN\propto\Phi(L,a)L^{3/2}\,dL\,da,
\end{equation}
where $\Phi(L,a)\,dL\,da$ is the space density of sources with
luminosity in the range $L\rightarrow L+dL$ and spins in the range
$a\rightarrow a+da$.  Transforming into the $(\dot{M},a)$-plane gives,
\begin{equation}
dN\propto \Phi(\dot{M},a)\eta^{3/2}\,d\dot{M}\,da.
\label{eqn:dndmda}
\end{equation}
Using our assumption that the mass accretion rate is independent of
spin, we can separate $\Phi(\dot{M},a)$ into an accretion rate
dependent space density $n(\dot{M})$ and a spin distribution function
$f(a)$, $\Phi(\dot{M},a)=n(\dot{M})f(a)$.   We can then integrate
Eqn.~\ref{eqn:dndmda} over $\dot{M}$ in order to determine the number
of sources in a flux-limited sample with spins in the range
$a\rightarrow a+da$:
\begin{equation}
dN_a\propto f(a)\eta(a)^{3/2}\,da\,\left(\int n(\dot{M})\,d\dot{M}\right).
\label{eqn:dnda}
\end{equation}
For illustration purposes, let us examine eqn.~\ref{eqn:dnda} in the
case of a completely flat spin distribution where $f(a)={\rm
  constant}$ for $a\in [0,a_{\rm max}]$ and is zero otherwise.  Thus,
half of the parent population as a whole has $a>a_{\rm max}/2$.  We
find that if $a_{\rm max}=0.95$, then half of sources in the flux
limited sample will have $a>0.67$; for $a_{\rm max}=0.99$ we find that
half of the sources in the sample have $a>0.73$.

Generalizing away from a flat spin distribution, we can consider spin
distribution functions of the form $f(a)\propto a^p$.   Within
this simple framework, we require $f(a)\propto a$ (i.e. $p=1.0$) 
in order to produce flux-limited samples where 
half of sources have $a>0.84$ (assuming $a_{\rm max}=0.95$).  For
high-spin weighted distribution functions such as this, the selection
bias is stronger; only 20\% of objects in the volume-limited parent
sample actually have $a>0.84$.  Of course, given the small number
statistics and highly inhomogeneous selection functions for the
current spin measurements, it is too early to draw any conclusions
about the need for a high-spin biased distribution function.

We are extremely grateful to our NASA and JAXA colleagues in the {\it
  Suzaku} project for enabling these Key Project data to be collected.
We thank Martin Elvis and Cole Miller for insightful conversations
throughout the course of this work, and the anonymous referee, who
  provided useful feedback that has improved this manuscript.  
This work was supported by NASA
under the {\it Suzaku} Guest Observer grant NNX09AV43G.

\bigskip


\bibliographystyle{apj}
\bibliography{adsrefs}

\end{document}